# Generalizable Deep Learning Approach for 3D Particle Imaging using Holographic Microscopy (HM)

SHYAM KUMAR M[1] AND JIARONG HONG[1,2,*]

[1]*University of Minnesota, Department of Mechanical Engineering, Minneapolis, MN, USA*
[2]*Saint Anthony Falls Laboratory, University of Minnesota, Minneapolis, MN, USA*
*\*jhong@umn.edu*

**Abstract:** Despite its potential for label-free particle diagnostics, holographic microscopy is limited by specialized processing methods that struggle to generalize across diverse settings. We introduce a deep learning architecture leveraging human perception of longitudinal variation of diffracted patterns of particles, which enables highly generalizable analysis of 3D particle information with orders of magnitude improvement in processing speed. Trained with minimal synthetic and real holograms of simple particles, our method demonstrates exceptional performance on various challenging cases including those with high particle concentrations and noises and a wide range of particle sizes, complex shapes, and optical properties exceeding the diversity of training datasets.

## 1. Introduction

Holographic Microscopy (HM) serves as a powerful technology for particle diagnostics across a spectrum of scientific and industrial applications [1-3]. It finds critical use in fields such as medicine and biology [4-6], manufacturing [7], environmental science [8,9], agriculture [10,11], and fundamental science [12-13]. In HM, holograms are formed by the interference of the light scattered by the particle field and unscattered portion of the same coherent illumination light [3]. In contrast to the brightfield imaging that captures only intensity, a hologram contains both phase and intensity data. This additional phase information facilitates a more comprehensive analysis of particle fields, revealing details about their morphology, optical, and biochemical properties. Using numerical reconstruction, one can obtain a three-dimensional (3D) particle field from the recorded two-dimensional (2D) holograms [3]. Moreover, HM's high resolution, extensive depth of field, cost-effectiveness, and compactness not only underscore its efficiency but also its potential for widespread adoption in analyzing diverse range of particles under natural and industrial settings.

Despite the transformative potential of HM, significant challenges hinder its broad application, particularly in real-time environments. The primary challenge lies in hologram processing, specifically, converting a 2D hologram into a comprehensive 3D particle field. Conventional processing typically involves three main steps [14-17]. First, pixels likely to contain individual particles are identified from either the extended focus image or the minimum intensity projection from the numerically reconstructed optical field, known as the reconstruction stack. Subsequently, focusing techniques, employing metrics such as Sobel-based filters or edge sharpness, ascertain the precise focal planes of particles within this stack. Lastly, segmentation methods are employed to extract particle morphology. Although thresholding is commonly applied for initial pixel identification and final segmentation, its efficacy varies across different particle types and often necessitates manual oversight [1,18,19], which complicates the adoption of HM in dynamic, real-world applications. The inherent complexities of conventional hologram processing [2,4,20] primarily restrict its use to academic settings, highlighting an urgent need for advancements in real-time processing capabilities that can adapt to varied scenarios. Deep learning (DL) has emerged as a promising tool for overcoming the inherent limitations of conventional hologram processing, significantly enhancing processing tasks such as reconstruction, image classification, feature segmentation, and phase recovery [21,22]. Unlike conventional methods, DL streamlines the feature extraction process by learning hierarchical and complex patterns from images, thereby obviating the need for manual hyperparameter tuning [21,23]. Furthermore, it supports both real-time and batch processing, broadening the applicability of holography beyond its conventional academic confines [24]. DL models for particle analysis in holography can generally be divided into two types: multi-stage and single-stage networks. Multi-stage networks employ separate DL models for particle detection and focal plane estimation. The YOLO (You Only Look Once) network is the most commonly used for particle detection [25-28], though support vector machines (SVM) [29] are also utilized for improved classification accuracy. For focal plane estimation, either classification or regression models may be applied. Classification models use convolutional neural networks (CNNs) to assign hologram images to discrete focus levels, thereby identifying the in-focus depth of particles [30-32]. In scenarios requiring continuous depth estimation, regression models are more suitable [25,33-35]. Most multi-stage models are effective for holograms with low particle concentrations, where individual particle fringes do not overlap. Single-stage models, in contrast, integrate particle detection, focal plane estimation, and segmentation within a single neural network. For example, the OSNet model outputs 3D particle coordinates directly from 2D holograms [36]. This model offers improved depth estimation and processing speed but can struggle with accuracy in high particle concentration scenarios, where depth estimation errors and boundary truncation affect particle detection performance [36]. To cope with these issues, U-Net architectures have been employed effectively [37-39], with modifications that include added layers and custom loss functions to boost particle extraction rates significantly over earlier models. Despite these improvements, U-Net models require extensive labeled datasets with a broad range and high diversity for effective training, which can limit their generalizability across varying holographic conditions, such as particle number densities, optical settings, and morphological diversity. For instance, a U-Net model trained on high-concentration particle fields may underperform on low-concentration fields, and vice versa [10]. Addressing this limitation often involves expanding the dataset with more diverse data, which further increases the size and complexity of the training set required for generalizability. In an innovative approach to enhance model adaptability, Chen et al. (2021) [40] introduced a physics-informed network that reconstructs 3D particle fields from 2D holograms by learning parameters that define the point spread function (PSF). While this model shows promise in terms of generalizability, it still faces challenges at high particle concentrations and is restricted by memory constraints when processing larger hologram sizes. The previous studies on DL for hologram processing are summarized in Table 1.

Despite the advancements in DL, many models developed thus far suffer from a lack of generalizability, with their effectiveness heavily influenced by factors like particle concentration, morphology, size ranges, and optical properties. This often results in inaccurate predictions that necessitate further post-processing, thus increasing the overall processing time. Efforts to enhance generalizability frequently result in compromises on model performance, potentially impeding the real-time processing of holograms in diverse applications such as product quality control in manufacturing, environment monitoring, and medical diagnostics [41-44]. In response to these challenges, our current work introduces a novel DL model designed to robustly process holograms across a wide spectrum of particle concentrations, shapes, and polydispersity. To achieve this, our model architecture is both physics-informed and tailored to prevent overfitting, enhancing its generalizability. Section 2 details the architecture of our proposed model; Section 3 evaluates the model's performance across various hologram types; Section 4 compares the model with other models and discusses its limitations; and Section 5 concludes with a summary and the implications of our findings.

**Table.1.** Summary of different DL models used for particle analysis in holography

| | | DL Model | Key Performance Indicators (KPI) | Limitations | References |
|---|---|---|---|---|---|
| Multi-stage network | Particle detection | YOLO Framework | - Real-time processing<br>- High detection accuracy | - Struggles with high particle concentrations and overlapping diffraction patterns | [25-28] |
| | | Support Vector Machines (SVMs) | - Classification accuracy | - Sensitive to noise<br>- Difficulty with overlapping patterns | [29] |
| | Particle localization | Classification models | - Depth estimation precision<br>- Rapid classification speed | - Performance decreases with high particle concentrations | [30-32] |
| | | Regression models | - Prediction accuracy<br>- Real-time capability | - Reduced localization accuracy in dense particle scenarios<br>- Challenges with boundary images | [25,33-35] |
| Single stage network | | OSNet | - Processing speed<br>- Accurate 3D coordinate output | - Performance may degrade with high particle concentrations or significant noise | [36] |
| | | U-Net-Based Models | - Accuracy in high concentration scenarios | - Purely data driven<br>- Less generalizable | [37-39] |
| | | MB-HoloNet | Generalizability<br>- Reduced reliance on labeled datasets | - Early development stage<br>- Challenges with scalability and practical implementation | [40] |

## 2. Methodology

A novel DL-based model is introduced to leverage the inherent 3D nature of diffraction patterns—specifically, the fringes around each particle (Fig. 1)—which are generated by the interference between the wave scattered from the particle and a reference wave. As illustrated in Fig. 1(a), this diffraction pattern converges towards the morphology of a particle at its in-focus plane ($z = 0$) along the longitudinal optical axis ($z$-axis) and diverges as it moves away from this plane. This pattern of convergence and divergence along the longitudinal axis is consistent regardless of the particle's morphology and optical properties, as demonstrated in Figs. 1(b) and 1(c). To date, DL models for particle hologram processing have primarily analyzed 2D lateral ($x$-$y$ plane) patterns. However, these patterns vary significantly across particles with different characteristics such as size and optical refractive indices and are vulnerable to cross-interference noise. This type of noise frequently occurs in scenarios with high particle concentrations where patterns from different particles overlap laterally, or with complex-shaped particles where diffraction patterns from different parts of the same particle overlap. In contrast, the longitudinal variation in diffraction patterns is inherently more robust against such cross-interference noises. Therefore, a DL model that leverages this 3D variation can significantly enhance adaptability and generalizability across diverse particle fields, effectively accommodating variations in particle concentration, morphology, and optical properties.

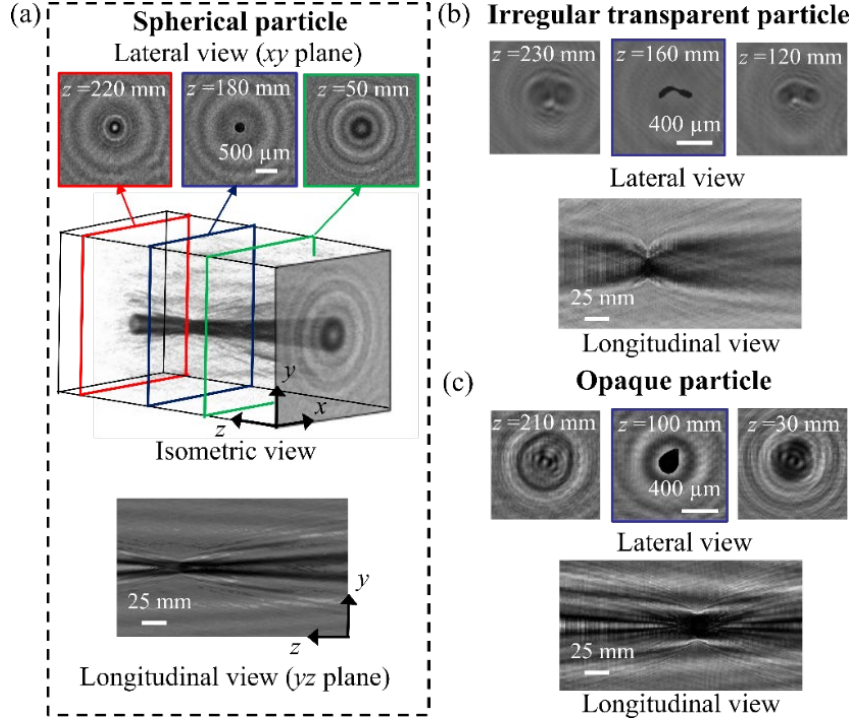

**Fig.1.** (a) 3D diffraction patterns generated from coherent light scattered from a spherical particle along the longitudinal direction. The top panel shows the cross section (xy plane, lateral view) of the diffraction patterns at three locations—before, at, and after the in-focus plane—represented by green, blue, and red bounding boxes, respectively, and the lower panel presents the longitudinal view (yz plane) of the pattern. The lateral and longitudinal views of 3D diffraction patterns for (b) transparent and (c) opaque irregular particle samples.

Consequently, a customized DL model is developed to analyze the 3D diffraction patterns of a particle hologram and extract the 3D location of each particle and its morphology. As illustrated in Fig. 2, the model input is a sequence of reconstructed planes from a hologram at different longitudinal positions with equal distance Δ apart between the neighbouring planes. The reconstruction can be conducted using different diffraction formulas such as the one below [3]:

$$u_p(x, y, z) = FFT^{-1}\left[FFT(I(x,y)) \times FFT\left(\frac{exp(jkz)}{jkz} exp\left\{j\frac{k}{2z}[(x^2 + y^2)]\right\}\right)\right] \quad (1)$$

Where $I(x, y)$ represents the original hologram, $u_p(x, y, z)$ denotes the 3D complex optical field, *FFT* stands for the Fast Fourier transform operation, and $k = 2\pi/\lambda$ and $\lambda$ correspond to the wavenumber and wavelength, respectively. The input information is analyzed simultaneously by two branches of the model, i.e., lateral and longitudinal variation networks, with the former focusing on the lateral variation and the latter aiming to analyze the longitudinal variation of diffraction patterns, respectively. The outputs of these two networks are subsequently merged and condensed through another customized network to obtain 3D particle position and size.

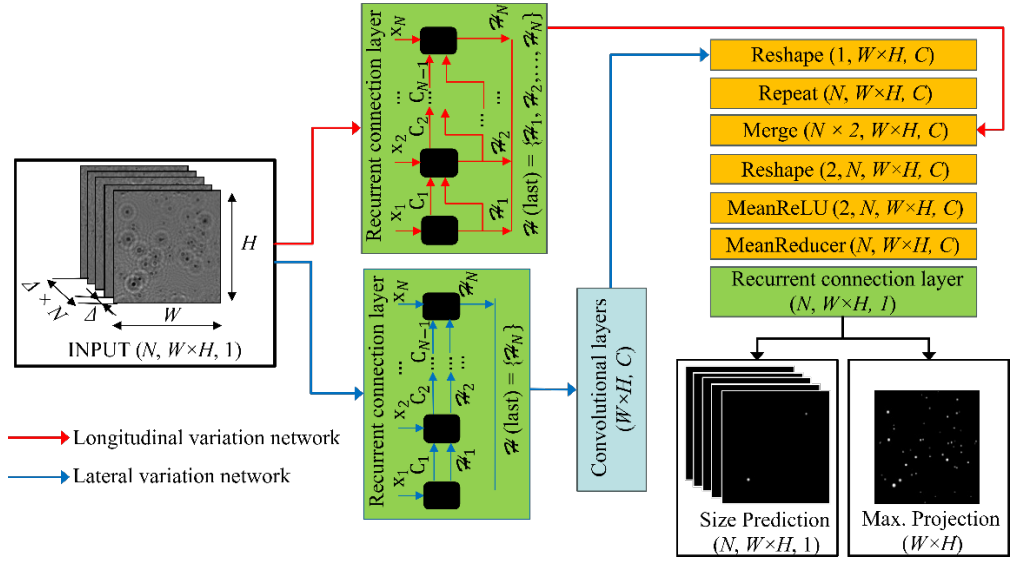

**Fig.2.** Schematics illustrating the network architecture of our deep learning model for hologram processing.

The lateral variation network uses a convolutional neural network (CNN) structure to extract information primarily associated with particle morphology. The series of convolutional layers consists of multiple layers with varying filter sizes and kernel sizes, each followed by a ReLU activation function (Fig. 2). This network begins with a recurrent connection layer, and its output retains only the hidden state $\mathcal{H}$ of the last image, denoted as $\mathcal{H}$ (last) = $\mathcal{H}_N$. This design allows summarizing the entire sequence into a single hidden state. The CNN architecture has thirteen consecutive Conv2D layers, each utilizing 3×3 kernels and ReLU activations. The network begins with two layers of 32 filters to capture low-level features, followed by two layers of 16 filters for refinement. It then includes four layers of 8 filters to detect intricate patterns and smaller structures within the hologram. To enhance feature integration and reconstruction, the number of filters progressively increases back to 16 and 32, culminating in a final Conv2D layer with 64 filters to consolidate the extracted features and accurately represent particle shapes. This hierarchical arrangement of decreasing and then increasing filter sizes enables the network to learn both detailed and abstract features, improving its ability to generalize across diverse particle sizes and morphologies. The structured depth and filter progression ensure robust feature extraction, thereby enhancing the accuracy and reliability of particle detection in complex holographic data. The output of the lateral variation network is then transformed through a reshape and repeat layer to match the dimension required for merging with the output of the longitudinal variation network in the subsequent process. While this output may contain some information regarding longitudinal variations, the series of convolutional layers followed by this step could potentially impact the model's ability to capture subtle variations across multiple consecutive images during auto-regressive inference [45]. This limitation necessitates the introduction of the second network, i.e., longitudinal variation network.

The longitudinal variation network is a sequential neural network designed to retain rich longitudinal information, addressing the limitations of the lateral variation network. This network is characterized by its recurrent connections, enabling it to capture dependencies between consecutive $N$ images along the longitudinal direction (Fig. 2). Structurally, the network incorporates a layer that resembles the long short-term memory (LSTM) module. However, standard LSTM fails to capture spatial correlations between input sequences due to fully connected layers treating each input independently. The proposed model aims to exploit spatial information, specifically convergence and divergence of diffraction patterns along the longitudinal direction (Fig. 1), which standard LSTM modules do not adequately address. Therefore, the longitudinal variation network employs a module that utilizes convolutional layers instead of fully connected layers to effectively capture spatial correlations along the longitudinal direction. In this context, the time sequence of images in recurrent connections is analogous to the sequence of images at different longitudinal reconstructed planes. Each image in the input sequence, from 1 to $N$, is processed by this layer, which outputs a corresponding hidden state $\mathcal{H}$ for each image, denoted as $\mathcal{H}$ (last) = $\{\mathcal{H}_1, \mathcal{H}_2, \ldots, \mathcal{H}_N\}$ (Fig. 2). This design allows the network to retain and propagate information along the longitudinal axis, facilitating a comprehensive analysis of pattern evolution. By capturing interactions between consecutive images, it ensures that information from previous images influences the processing of subsequent ones, which is crucial for detecting the in-focus plane and analyzing longitudinal pattern variations.

The outputs of the longitudinal and lateral variation network are merged and fed into another customized sequential network featuring with two transformation layers: MeanReLU and MeanReducer (Fig. 2). The objective of these layers is to enhance feature representation and reduce dimensionality, respectively. Specifically, MeanReLU computes the mean longitudinally and applies ReLU separately to the positive and negative components of the tensor, allowing it to distinguish between positive activations, indicating feature presence, and negative ones, representing absence or noise. This significantly improves model accuracy by offering detailed insights into features and noise characteristics, effectively reducing false positives. Meanwhile, MeanReducer computes the mean along the second axis of the tensor, effectively collapsing it while preserving the other dimensions. This dimensionality reduction aims to aggregate information across the branches, refining the feature representation in preparation for the subsequent layer. The final output of the model consists of two channels: size prediction and the corresponding maximum projection (Fig. 2). The input and output size prediction channels maintain the same dimensions, with each prediction in the stack corresponding to the subsampled diffraction pattern of the input. Particles are predicted as white labels with corresponding morphologies on their respective in-focus planes against a black background. The maximum projection output channel indirectly assists the size prediction channel in training by acting as a regularization mechanism, helping to learn the most relevant features across the longitudinal direction.

To ensure precision in both location and shape detection, we employ customized loss functions specifically designed for this model. As detailed in Eq. 2, our approach incorporates three distinct loss functions tailored to enhance performance. In this context, $N$ denotes the total number of subsampled images, while $W$ and $H$ represent the image width and height, respectively. The symbols $X_{i,j}$ and $Y_{i,j}$ refer to the true and predicted data, for the $j^{th}$ pixel in the $i^{th}$ sample, respectively, while $X_i$ and $Y_i$ indicate the true and predicted data for the entire $i^{th}$ sample. The weights are represented by $\delta_i$ and $\gamma_{i,j}$. In holograms, detecting smaller particles is challenging due to weak signals, while larger particles require thresholding to accurately delineate their shapes, with threshold values varying across cases and potentially limiting generalizability. Consequently, the development of the loss function focused on two primary goals: (1) accurately detecting smaller particles in holograms and (2) generating precise intensity maps for each particle in the predictions. The first term represents the plane-weighted mean square error (MSE), where $\delta_i$ is set to 5.0 when there is at least one label present, and 1.0 if there are no labels. While plane-weighted MSE is primarily intended to improve longitudinal location accuracy, it also enhances the detection of smaller droplets. This effect can be attributed to the extended interference signal of a particle along the longitudinal direction compared to the lateral direction in holography. The second loss function employed is binary cross-entropy (BCE), which has been shown to be efficient in image segmentation tasks [46] and helps to achieve a more uniform threshold across particles. However, to address the class imbalance due to the relatively smaller number of white pixels ($X_{i,j} = 1$) compared to black pixels ($X_{i,j} = 0$), we also utilize a class-weighted MSE. In this function, $\gamma_{i,j}$ is set to 2.0 for $X_{i,j} = 1$ and 1.0 for $X_{i,j} = 0$. The optimal weight for each class is determined from the class frequencies in the training dataset. Specifically, the inverse of the average class frequencies, calculated from the maximum projections in the second output of the architecture (as shown in Fig. 2), across the training dataset, is used to estimate the class weights. This adjustment is crucial for improving the model's ability to accurately segment and identify particle shapes despite the imbalance. Although varying class weights can influence predictions, our analysis shows that the selected weight achieves a balance that effectively addresses class imbalances while maintaining generalizability. This enables consistent model performance across different cases without needing re-optimization for each new scenario. Since class weight affects how the model penalizes misclassifications in specific regions, an optimal weight allows the model to focus on particle features across various scales, supporting performance across particle types. The total loss is averaged over the batch size to ensure stable training. The combination of these loss functions, each addressing different aspects of the detection and segmentation tasks, underlines the tailored nature of our approach. This specialized design helps in achieving higher accuracy and robustness compared to traditional methods.

$$L = \frac{1}{N}\sum_{i=1}^{N}\left[(\delta_i \cdot (X_i - Y_i)^2) + BCE\,(X_i, Y_i) + \left(\frac{1}{W \times H}\sum_{j=1}^{W \times H}\gamma_{i,j} \cdot (X_{i,j} - Y_{i,j})^2\right)\right] \quad (2)$$

The training dataset consists of reconstructed planes with corresponding training targets. For the training targets, binary images containing only in-focus particles with their true shapes in white on their respective in-focus planes against a black background are used for size prediction. The maximum projection of these binary images is used as the second training target (Fig. 2). Both synthetic and real holograms are used for training. The robustness of the proposed architecture in detecting particle fields is tested using three models trained with synthetic holograms of three different types of particle fields: tracers, polydispersity, and irregular transparent and opaque particles. To demonstrate the generalizability of this approach for processing experimentally obtained real holograms across varying particle concentrations, optical properties, and shapes, the model, initially trained with synthetic data, is fine-tuned using limited experimental data and then tested on holograms of entirely different particle fields.

## 3. Result and discussion

In this section, we demonstrate the capability of our approach in analyzing particle holograms across a wide range of particle concentrations, sizes, morphologies, and optical properties. Specifically, using synthetic holograms, we showcase our approach is able to analyze dense particle holograms with number concentrations exceeding the highest levels presented in the literature, with large dynamic size range as well as with various particle morphology and optical properties, in sections 3.1 to 3.3, respectively. In section 3.4, we further demonstrate our approach can be generalized to analyze real holograms with a variety of particle properties (i.e., concentrations, morphology, and optical properties).

### *3.1 Synthetic holograms: tracer particles*

To demonstrate our approach for a wide range of particle concentrations, synthetic holograms with particle concentrations ranging from $1\times10^{-3}$ to $1.5\times10^{-1}$ particles per pixel (ppp), corresponding to approximately 66 to 9830 particles per hologram, are used. In total 200 synthetic holograms of size $256\times256$ pixel$^2$ for tracer-like particles of size 1 pixel are generated using Rayleigh-Sommerfeld diffraction equation [3] for the training. Sample holograms for three number concentrations are shown in the top panel of Fig. 3. Overlap of diffraction patterns of individual particles occurs with an increase in concentration from $1\times10^{-3}$ (Fig. 3a) to $1\times10^{-2}$ (Fig. 3b). This overlapping becomes more evident at high number concentrations (Fig. 3c). For a wavelength of 632 nm and resolution of 10 μm/pixel, each hologram is reconstructed for a longitudinal range from $z = 1000$ μm to $z = 2280$ μm, with $\Delta$ of 10 μm. For each training batch, a training and validation generator is used to choose $N = 32$ consecutive reconstruction images that correspond to the same hologram.

The proposed approach demonstrates a high extraction rate (ER), minimal lateral errors, and relatively smaller longitudinal errors even at higher particle concentrations compared to previous studies. This is evidenced in Fig. 3(d), where the test results for particle concentrations from $1\times10^{-3}$ to $1.5\times10^{-1}$ ppp, with each particle concentration tested using 100 holograms generated with the same parameters as those used for the training case, are shown. Interestingly, ER remains above 90% for concentrations less than $1.1\times10^{-1}$ ppp, which is approximately twice the maximum concentration used in Shao et al. (2020) [39]. Similar to Shao et al. (2020) [39], in this particle concentration range, longitudinal mean error $\Delta z$ remains less than 4 voxels. The performance remains robust even at 2.5 times the concentration used in previous DL-based approaches [39], with ER above 88%. It is noteworthy that the mean lateral errors, $\Delta x$ and $\Delta y$, remain less than 1 voxels for all particle concentrations.

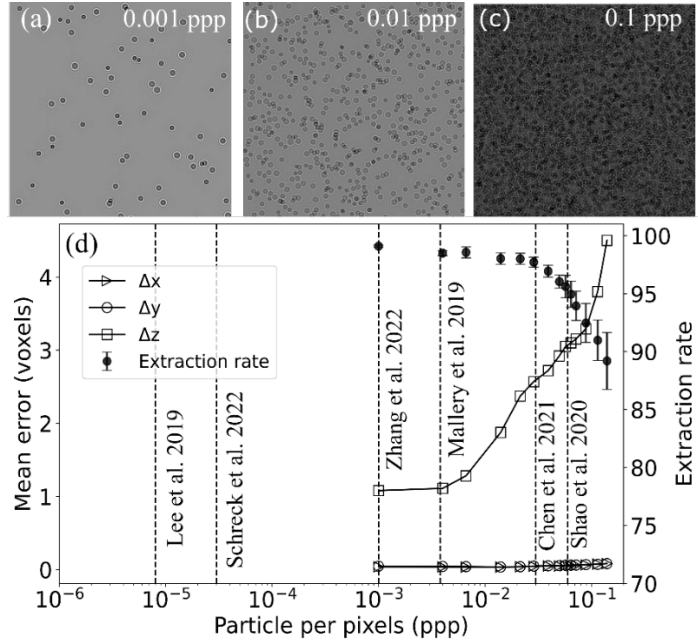

**Fig.3.** (a)-(c) Sample holograms for different particle number concentrations. (d) Variation of mean error (first $y$-axis) for lateral $\Delta x$, $\Delta y$, and longitudinal $\Delta z$ directions for different particle concentrations. Corresponding variation in extraction rate is shown on the second $y$-axis. The variation in extraction rate for a given bin of particle concentration is plotted as the standard deviation. Particle concentrations used in previous DL-based hologram processing studies are shown using vertical dashed lines.

*3.2 Synthetic holograms: varying size distribution*

The performance of our approach is assessed across larger dynamic size ranges using synthetic holograms of a field of polydisperse water droplets. Despite holograms with increased cross-interference, the proposed approach consistently achieves ER exceeding 95% and maintains false positive rates (FPR) below 5%. Model is trained using 150 holograms with particle concentration ranging from $1.9\times10^{-4}$ to $1.9\times10^{-3}$ ppp, distributed over a z-depth of 25 cm with z-step of 1000 μm. The particle diameters range from 1 to 20 pixels, corresponding to 18.2 μm to 364 μm. Holograms are reconstructed at a wavelength of 632 nm, and training utilizes 25 consecutive images ($N = 25$) corresponding to the same hologram per batch. The performance of the trained model is evaluated using 250 synthetic holograms featuring the same particle field as the trained data.

As evident in Fig. 4, the predicted contours of particles (yellow contour) match very well with the ground truth (blue filled circles) for particles of a wide range of size distributions. The total ER and FPR across the 250 test synthetic holograms are estimated to be 97% and 3.3%, respectively. Additionally, over 95% of the detected droplets have a longitudinal error ($\Delta z$) that is less than the particle diameter, indicating accuracy on par with the best performance using both conventional [47-49] and deep learning-based [25-40] algorithms.

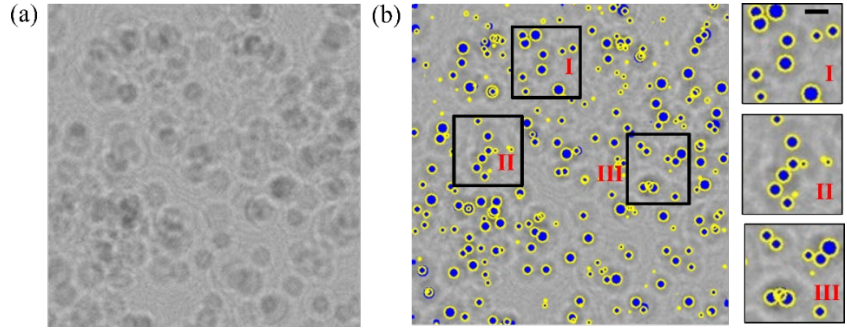

**Fig.4.** (a) Sample synthetic hologram of a polydisperse droplet field. (b) Prediction (in yellow contour) and ground truth (in blue) superimposed on the hologram. For clarity, zoomed views from three different regions of (b) are shown on the sides. The scale bar in all the zoomed view is equal to 500 μm.

### *3.3 Synthetic holograms: Varying particle shapes and optical properties*

To further assess the generalizability of our method, we evaluate its performance using synthetic holograms of particles with varying shapes and optical properties. Specifically, particles of varying shapes are generated from a random noisy image. A random number generator is used to create the noisy image, followed by Gaussian blur (with a kernel size of 8 pixels for this study) and thresholding to generate binary contours. These irregular binary contours serve as masks for synthetic hologram generation. Particle masks are generated as transparent, semi-transparent, or opaque based on the imaginary part of their index of refraction (IoR). Transparent particles have an imaginary part less than 0.01, opaque particles have an imaginary part greater than or equal to 0.1, and semi-transparent particles fall in between. The reflectance and absorption are estimated using Fresnel equations and the imaginary part of the specified IoR based on the required optical properties, with transmittance determined by considering both. These values, representing the portion of light blocked by the particles, are used to generate the holograms. As shown in Fig. 5, our model, trained using holograms containing solely transparent particles, successfully produces accurate contours for both transparent and opaque particles of complex shapes. Notably, these complex shapes are distributed far from the focus plane, resulting in noisy holograms due to strong cross-interference, yet our approach still performs robustly.

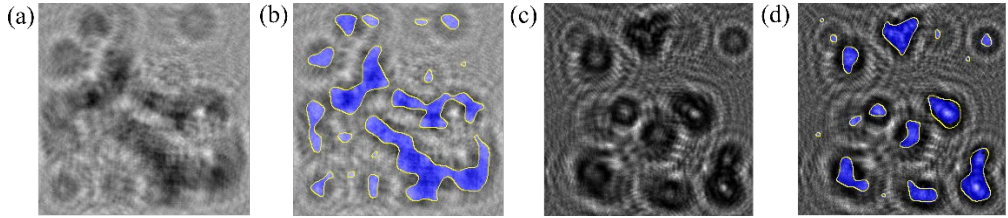

**Fig.5.** (a) Sample synthetic hologram (b) with the superimposed prediction (in yellow contour) and ground truth (in blue filled shape) for a transparent particle field. The corresponding images for the opaque particle field are shown in (c) and (d).

Qualitatively, based on the test of 2000 holograms, our approach achieves ER >95%, FPR <5%, and >0.9 intersection-over-union (IoU) for particles of varying shapes and optical properties. Specifically, the performance for transparent particles (i.e., 98% ER, 1.7% FPR, 0.94 IoU) is slightly better than for opaque particles (i.e., 96% ER, 3.8% FPR, 0.9 IoU). This difference can be attributed to the use of solely transparent particles for training and the loss of some information due to light being blocked by the opaque particles. Overall, our results imply that our approach is generalizable for particle fields containing particles of varying shapes and optical properties.

### *3.4 Real hologram: Varying particle concentration, optical property, and shape*

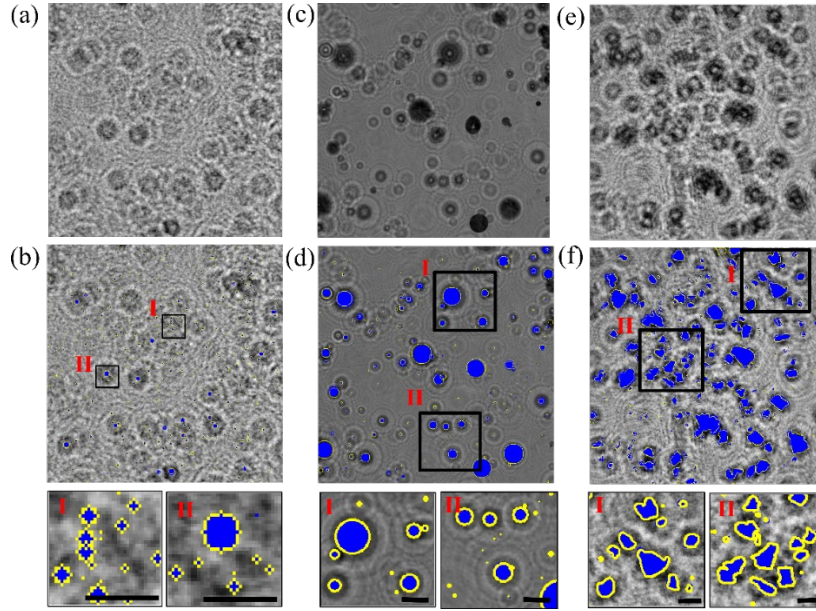

**Fig.6.** Sample hologram (first row) and corresponding superimposed prediction and ground truth (second row) for water spray droplet field (a and b), oil in water (c and d), and sugar particles (e and f). For clarity, two zoomed views of each prediction are shown on the bottom panel. The ground truth is shown in blue, and the predictions are shown in yellow. The scale bars in all the zoomed views are equal to 500 μm.

In the last three subsections (3.1–3.3), we used a synthetic dataset to test the model's performance. By utilizing a larger synthetic dataset, we eliminate the need for time-consuming manual labeling required to obtain ground truth in experimental holograms. This approach allows us to generate a diverse, extensive dataset for testing, which is challenging to achieve experimentally. Here, we demonstrate the generalizability of our approach for processing experimentally obtained real holograms across varying particle concentrations, optical properties, and shapes. To address the difference in diffraction patterns in real holograms arising from phase differences due to particle depth, which do not appear in synthetic holograms generated using 2D particle masks, our model, initially trained with synthetic data, is fine-tuned with limited experimental data to enhance its generalizability. Specifically, the model trained using synthetic holograms of a polydisperse droplet field (section 3.2) is fine-tuned using real holograms of spherical water droplets with very low number concentrations (~$1.0 \times 10^{-3}$ ppp) [10]. For the fine-tuning, only 20% of the training data used for synthetic holograms, i.e., 30 holograms instead of 150, is utilized, significantly reducing the manual labelling effort. The fine-tuned model is tested on three distinct and complex real holograms: (i) spherical water droplets with high number concentrations (~$3.0\times10^{-3}$ ppp) and strong cross-interference, (ii) oil in water with a larger size range and different optical properties, and (iii) irregular sugar particles. These datasets present real-world challenges, including cross-interference from overlapping particles, substantial noise, and artifacts arising from varying optical properties and uncontrolled illumination. The spray nozzle used for testing is the TP11001, which, according to the ASABE standard [50], produces very fine droplets dispersed approximately 40 cm longitudinally, resulting in noisy holograms with extensive diffraction pattern overlap. For oil-in-water cases, in addition to higher dynamic ranges compared to the trained model, changes in optical properties occur, particularly near the interface between water and oil droplets. Since the model was trained on water droplets suspended in air, the altered optical properties encountered in oil-in-water scenarios present additional challenges for accurate detection and characterization. Although the model was trained exclusively on spherical droplets, we evaluate its performance on particles of varying shapes. Given these complexities, processing these holograms poses difficulties for both conventional techniques and other deep learning models.

Qualitatively, using the same fine-tuned model, a visually evident and promising overlap between the predictions and ground truth is observed across all three distinct holograms (Fig. 6), demonstrating the generalizability and robustness of our approach. Specifically, our approach can detect particles with very high cross-interference, such as individual droplets that are very close to each other and smaller droplets in close vicinity of larger droplets, as shown in the zoomed views I and II, respectively, for the water droplets case in Fig. 6. Notably, while the proposed model is trained for water droplets with a maximum size of 20 pixels, it could detect particles as large as four times the trained maximum size and particles of different optical properties, i.e., oil in water (Fig. 6, second column). Furthermore, the model, solely trained with spherical shapes, has shown a high overlap between prediction and ground truth for the irregular sugar particle field case (Fig. 6, third column). Quantitatively, high ER and low FPR are observed for the high concentration case (94% ER, 1.3% FPR) and for the larger size range and optical property case (92% ER, 4.8% FPR). An IoU of ~0.94 for the prediction and ground truth of the irregular sugar particle case demonstrates that the model effectively interpolates the complex combination of the spherical scattered wave and planar reference wave learned in the spherical water droplet case to the irregular sugar particle case. Here, using a smaller training dataset—20% of that used for synthetic holograms—and demonstrating effective performance on cases (such as irregular shapes, varying dynamic ranges, etc.) not represented in the training set highlights the strong generalizability of the proposed model and reduces the need for an extensive dataset.

To further demonstrate the capability of our method, the proposed model, without any additional training, is used to analyze more challenging holograms with increased droplet concentration and greater longitudinal distribution. Specifically, we used a hologram generated with a hollow-cone nozzle (TXA8003 VK). The hologram in Fig. 6(a), produced by the TP11001 flat spray nozzle, represents the noisiest condition among flat spray nozzles. In contrast, the hollow-cone nozzle exhibits a significantly higher droplet concentration and a broader longitudinal distribution. Specifically, the TXA8003 VK nozzle has more than double the concentration (~$7.0\times10^{-3}$ ppp) and distributes droplets over 60 cm, compared to 40 cm for the TP11001. This higher concentration and extended spray span increase noise in the hologram, as seen in the comparison between Fig. 6(a) and Fig. 7(a). The detection using the proposed model across the entire depth of the spray is shown in Fig. 7(b). 200 reconstruction planes were generated and analyzed manually to establish the ground truth. The corresponding region is marked with a red box in Fig. 7(b). Within this region, the extracted true particles and the unpaired particles are identified (Fig. 7c), yielding an ER of 93% and an FPR of 5.8%, demonstrating the model's robustness under high-noise conditions.

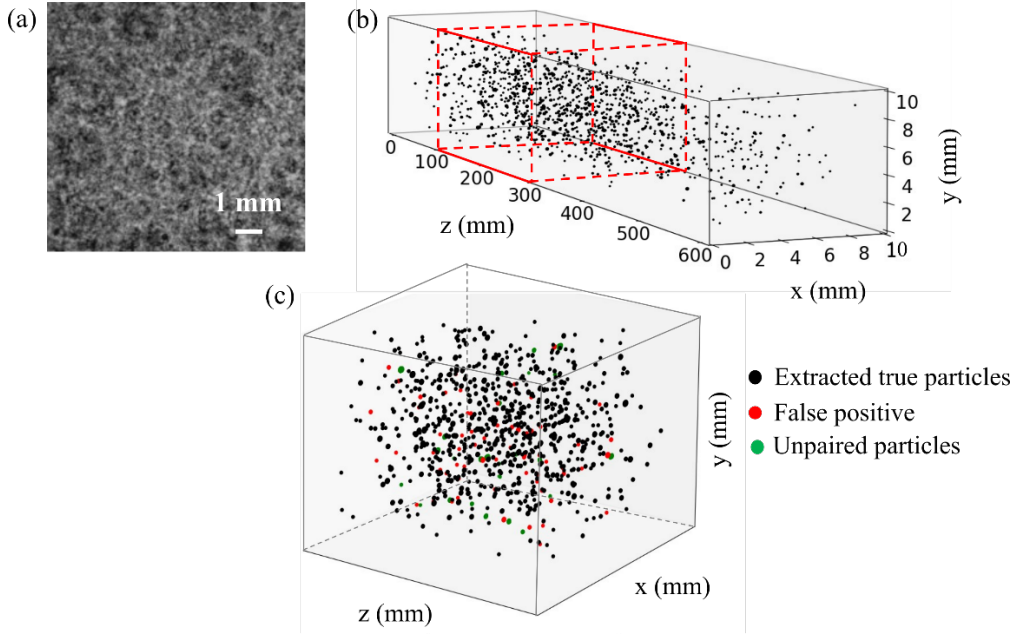

**Fig. 7.** Experiment dataset using TXA8003 VK hollow-cone spray nozzle. (a) sample hologram (b) corresponding detection (c) zoomed in view of the region marked in red in (b). The black dots are the extracted true particles, red dots are false positive and green dots are the unpaired particle from the ground truth.

### 4. Model performance comparison and Model limitations

In this section, we compare the performance of the proposed model with other DL models in terms of accuracy and speed. The limitations of the proposed model are also discussed. As described in Section 3.1, the proposed model achieved an ER exceeding 90% for particle concentrations twice as high as those used in the previous U-Net model [39]. The lateral and longitudinal errors of the proposed model are comparable to those of the U-Net model. For the spray case, the same U-Net model proved more effective than conventional non-DL methods [39]. However, droplet ER decline in high-density, polydisperse droplet fields due to an increase in false negatives. Conversely, for low-density sprays, FPR increases due to class imbalance. To address this, Kumar et al. (2023) [10] used a modified U-Net alongside a VGG16 classifier. For very fine sprays, the performance of U-Net + VGG16 (94.8% ER, 2.8% FPR) is comparable to that of the proposed model (94% ER, 1.3% FPR). In summary, the proposed model can be applied across different cases, consistently delivering comparable or improved performance without the need for separate models (U-Net and U-Net + VGG16).

One of the goals of using DL models for hologram processing is to achieve real-time performance. While previous models [25,37-40] have made promising strides, our proposed model significantly improves processing speed over conventional algorithms and U-Net-based models. For a 512×512 pixel hologram, conventional processing takes 3–4 minutes, U-Net and U-Net + VGG16 require approximately 30–40 seconds on an RTX 4090 GPU, whereas our model completes processing in just 4–6 seconds—nearly seven times faster. Further improvements in processing speed can be achieved by optimizing the number of floating-point operations (FLOPs), integrating transformer architectures, and reducing the number of layers.

Such improvements can enable real-time processing of holograms. Additionally, our model is much smaller (4.3 MB) compared to U-Net (376 MB) and U-Net + VGG16 (303 MB), enabling faster loading, lower latency, and reduced memory usage, making it ideal for deployment on resource-constrained or edge devices.

The model demonstrates robustness in predicting particle fields within complex holograms (Figs. 6 and 7), though there remains room for improvement. Currently, 3D particle imaging refers to detecting particle shapes and identifying a single in-focus plane. However, detecting particles with complex 3D shapes remains challenging. While the model accurately detects irregular particles (Sections 3.3 and 3.4) with small longitudinal extension relative to the reconstruction step size, $\Delta$ (Fig. 2), it struggles with structures like diatom chain [51] or spray ligaments, where longitudinal extension exceeds $\Delta$. In these cases, different parts come into focus on various planes, and the model—trained with 2D rather than 3D masks—cannot associate in-focus segments across planes. Future research will focus on using 3D masks to improve detection of such complex structures.

## 5. Conclusion

In this study, we introduced a novel deep learning (DL) model designed to leverage the inherent three-dimensional (3D) nature of diffraction patterns in holography for particle detection and characterization. Unlike existing DL models that analyze 2D lateral patterns, our model utilizes longitudinal variations in diffraction patterns, enhancing adaptability and generalizability across diverse particle fields. The model consists of two branches: a lateral network using convolutional neural networks for particle morphology and a longitudinal network employing a sequential neural network for longitudinal variations, improving detection accuracy for in-focus particles. Customized transformation layers enhance feature representation and reduce dimensionality, leading to precise particle morphology and location predictions resistant to the influence of prevalent background noises. With training under a limited amount of synthetic data, our model can be directly applied to holograms with particles of varying sizes, shapes, and optical properties beyond those in the training dataset. Our test has shown our model can achieve excellent performance (>90% ER, <5% FPR, >0.9 IoU) for even challenging cases including holograms of tracer particles with exceptionally high concentration and larger dynamic size ranges. Moreover, with little fine-tuning using a few simple real water droplet holograms, our model has further demonstrated its robustness across diverse particle fields, including higher concentration spherical droplets, oil in water, and irregular sugar particles. Overall, our DL-based approach significantly improves hologram processing, offering superior performance and generalizability compared to existing methods.

Our approach introduces fundamental changes to the deep learning architecture, departing from conventional methods and state-of-the-art networks. These innovations lead to notable improvements in generalizability, robustness, and processing speed. Such enhancements are crucial for bridging the gap between using HM as a research tool and deploying it as a commercially viable technology for real-time applications. This advancement facilitates the broader adoption of HM in various fields, including process monitoring in manufacturing, environmental surveillance, and point-of-care medical diagnostics. Additionally, our study offers valuable insights into deep learning-based computational imaging, inspiring novel network architectures that address the critical issue of generalizability, which has similarly limited the advancement of these technologies, including HM.

**Funding.**
**Acknowledgments.** The authors thank Dr. Chris Hogan from the University of Minnesota and Dr. Steven Fredericks from WinField United for the fruitful discussions.
**Disclosures.** The authors declare no conflicts of interest